\documentstyle[preprint,aps]{revtex}

\begin{document}

\draft
\title{A spin-cell for spin current}

\author{Qing-feng Sun$^1$, Hong Guo$^1$, and Jian Wang$^2$}

\address{$^1$Center for the Physics of Materials and Department
of Physics, McGill University, Montreal, PQ, Canada H3A 2T8.\\
$^2$Department of Physics, The University of Hong Kong, Pokfulam
Rood, Hong Kong, China}

\maketitle

\begin{abstract}
We propose and investigate a spin-cell device which provides the
necessary spin-motive force to drive a spin current for future
spintronic circuits. Our spin-cell have four basic
characteristics: (i) it has two poles so that a spin current flows
in from one pole and out from the other pole, this way a complete
spin-circuit can be established; (ii) it has a source of energy to
drive the spin current; (iii) it maintains spin coherence so that
a sizable spin current can be delivered; (iv) it drives a spin
current without a charge current. The proposed spin-cell for spin
current should be realizable using technologies presently
available.
\end{abstract}

Traditional electronics is based on the flow of charge: the spin
of electron is ignored. The emerging technology of spintronics
will make the leap such that the flow of spin---in addition to
charge, will be used for electronic applications\cite{ref1,ref2}.
A spin current is produced by the motion of spin-polarized
electrons, therefore spin current is typically associated with
spin-polarized charge current\cite{ref1}. Nevertheless, if one can
generate an ideal situation as shown in Fig.(1a) where spin-up
electrons move to the right while an equal number of spin-down
electrons move to the left, then there will be no net charge
current because $I_e\equiv e(I_{\uparrow}+I_{\downarrow})=0$ where
$eI_{\uparrow},eI_{\downarrow}$ are charge currents due to spin-up
and spin-down electrons, respectively. There will be, however, a
finite spin current: $I_s\equiv
\frac{\hbar}{2}(I_{\uparrow}-I_{\downarrow})$ where $\hbar$ is the
reduced Planck constant. Considering the interesting and important
future perspective of spin-current circuit, it is crucial to have
a spin-cell that satisfies the four characteristics discussed in
the Abstract and it produces the flow pattern of
Fig.(1a)\cite{ref3}. In this paper we theoretically propose and
analyze such a spin-cell.

Our spin-cell is schematically shown in Fig.(1b). It consists of a
double quantum-dot (QD) fabricated in two-dimensional electron gas
(2DEG) with split gate technology, and each QD is contacted by an
electrode. Note that no magnetic material is involved. The two QDs
and their associated contacts to the electrodes serve as the
``positive/negative'' poles of the spin-cell. The two electrodes
maintain the same electrochemical potential $\mu_L=\mu_R$ ({\it
i.e.} no bias voltage is applied on them). The size of the
spin-cell structure is assumed to be within the spin coherence
length which can be as long as many microns for 2DEG. We control
the QD energy levels by gate voltages $V_{g\alpha}$ where
$\alpha=L,R$ indicates the left/right QD. Both QD levels are
controlled by an overall gate voltage $V_g$, see gate arrangements
in Fig.(1b). In order to distinguish spin-up electrons from
spin-down electrons, a spatially {\it non-uniform} external
magnetic field $B_\alpha$ is applied to the two QDs--perpendicular
to the QD plane. An extreme case of non-uniformity is $B_R=-B_L$,
{\it i.e.} equal in value but opposite in direction. This
particular magnetic field distribution is not necessary at all for
the operation of our spin-cell, but it helps us to discuss its
physics. Finally, the energy source of our spin-cell is provided
by shining a microwave radiation with strengths
$\Delta_L/\Delta_R$ for the left/right QDs. Because, typically,
the microwave frequency is far less than the plasma frequency of
the material covering the QDs, the effect of the microwave field
is to induce a high frequency potential variation $\Delta_{L/R}
\cos \omega t$ in the left/right QD and their leads.\cite{ref4}
When $\Delta_L \not= \Delta_R$, a time-dependent potential
difference, $\Delta \cos \omega t =(\Delta_L -\Delta_R)\cos\omega
t$, exists between the two QDs. An a.c. electric field ${\bf
E}(t)$ in the middle barrier is therefore established due to the
microwave radiation (see Fig.(1c)). Then, electrons can absorb
photons when they pass the middle barrier of the device. The
establishment of ${\bf E}(t)$ across the two QDs is necessary for
our spin-cell to work, here we use a non-uniform microwave
radiation to achieve this effect as has already been carried out
experimentally\cite{ref7}, but other possibilities also exist.

Before we present theoretical and numerical results of the device
in Fig.(1b), we first discuss why it works as a spin-cell. The
physics is summarized in Fig.(1c). To be specific, let $B_R$ point
to $-z$ direction and $B_L$ to $+z$ direction. Due to Zeeman
effect, a spin-degenerate level $\epsilon_R$ on the right QD is
now split into spin-down/up levels
$\epsilon_{R\downarrow}<\epsilon_{R\uparrow}$. On the left QD, it
is $\epsilon_{L\uparrow}<\epsilon_{L\downarrow}$. Electrons in the
electrodes can now tunnel into the QD: on the right a spin-down
electron is easier to tunnel because level
$\epsilon_{R\downarrow}$ is lower, while a spin-up electron is
easier to tunnel into the left QD. Once levels
$\epsilon_{R\downarrow},\epsilon_{L\uparrow}$ are occupied, the
charging energies $U_R,U_L$ of the two QDs push the other two
levels $\epsilon_{R\uparrow},\epsilon_{L\downarrow}$ to higher
energies $\epsilon_{R\uparrow}+U_R,\epsilon_{L\downarrow}+U_L$,
and the energy level positions indicated by the solid horizontal
lines of Fig.(1c) are established. Next, the spin-down electron on
the right QD can absorb a photon and make a transition to the
level at $\epsilon_{L\downarrow}+U_L$ on the left QD: afterwards
it easily flows out to the left electrode because
$\epsilon_{L\downarrow}+U_L>\mu_L$. This process is indicated as
$A-$. Similarly the spin-up electron on the left QD flows out to
the right electrode after absorption of a photon, indicated by
$A+$. This way, driven by the potential variations of the QD
induced by the microwave field, a spin-down electron flows to the
left while a spin-up electron flows to the right of the spin-cell,
and the continuation of the $A-,A+$ processes generates a DC spin
current that flows from the left electrode, through the spin-cell,
and out to the right electrode. Clearly, if the two processes are
absolutely equivalent, there will be no charge current and only a
spin current. Finally, since the spin-motive force is provided by
a time-dependent change of the electronic potential landscape of
the QD, there is no spin flip mechanism and the spin current
flowing through the spin-cell is conserved, {\it i.e.}
$I_{s,L}=-I_{s,R}=I_s$. Our device then satisfies the four
characteristics of a spin-cell discussed in the Abstract.

The last paragraph discusses the operation principle of the
spin-cell for spin current, but there are other interesting device
details which can only be obtained by detailed theoretical and
numerical analysis for which we now turn. The spin-cell of
Fig.(1b) is described by the following
Hamiltonian:\cite{ref4,ref5}
\begin{eqnarray}
& H & = \sum\limits_{\alpha\sigma} \left[\epsilon_{\alpha}
+W_{\alpha}(t) -(1/2)\sigma g\mu B_{\alpha} \right]
d^{\dagger}_{\alpha\sigma} d_{\alpha\sigma}  \nonumber \\
& &  + \sum\limits_{\alpha} U_{\alpha} d^{\dagger}_{\alpha
\uparrow}  d_{\alpha \uparrow} d^{\dagger}_{\alpha \downarrow}
d_{\alpha \downarrow} + \sum\limits_{k\sigma\alpha} \left[
\epsilon_{\alpha k} +W_{\alpha}(t) \right]
a^{\dagger}_{\alpha k\sigma} a_{\alpha k\sigma} \nonumber \\
& & +  \sum\limits_{k\sigma\alpha} \left[ t_{\alpha k}
a^{\dagger}_{\alpha k\sigma} d_{\alpha\sigma} + H.c. \right]
+\sum\limits_{\sigma} \left[ t_C d^{\dagger}_{L\sigma} d_{R\sigma}
+H.c. \right] \label{hamiltonian}
\end{eqnarray}
where $a^{\dagger}_{\alpha k\sigma}$($a_{\alpha k\sigma}$) and
$d^{\dagger}_{\alpha\sigma}$($d_{\alpha\sigma}$) are creation
(annihilation) operators in the electrode $\alpha$ and the dot
$\alpha$, respectively. The left and right QD includes a single
energy level $\epsilon_{\alpha}$, but has spin index $\sigma$ and
intradot Coulomb interaction $U_{\alpha}$. To account for the
magnetic field $B$, the left/right QD's single particle energy has
a term $-(1/2)\sigma g\mu B_{\alpha}$, in which we have required a
different magnetic field strength for the two QDs, {\it i.e.} $B_L
\not=B_R$. $t_C$ and $\Gamma_{\alpha} \equiv 2\pi \sum_k
|t_{\alpha k}|^2 \delta(\epsilon-\epsilon_{\alpha k})$ describe
the coupling strength between the two QDs, and between electrode
$\alpha$ and its corresponding QD, respectively. The microwave
irradiation is given by\cite{ref4,ref5} $W_{\alpha}(t)
=\Delta_{\alpha} \cos\omega t$ and it produces an adiabatic change
for the single particle energy. Here we permit the microwave field
to irradiate the entire device including the electrodes, and we
require a difference in the radiation strength
$\Delta_L\not=\Delta_R$.

Our theoretical analysis of the spin-cell is based on standard
Keldysh nonequilibrium Green's function (NEGF)
theory\cite{ref4,ref5} which we briefly outline here. First, we
perform an unitary transformation of the Hamiltonian with an
unitary operator $U(t)= exp\left\{ i\int^t_0 dt'
\sum\limits_{\alpha} W_{\alpha}(t') \hat{D}\right\}$,
where
$\hat{D}\equiv \sum\limits_{k\sigma} a^{\dagger}_{\alpha k\sigma}
a_{\alpha k\sigma} +\sum\limits_{\sigma} d^{\dagger}_{\alpha
\sigma} d_{\alpha \sigma}$. The Hamiltonian $H$ is transformed to
the following form,
\begin{eqnarray}
H & = &
  \sum\limits_{\alpha\sigma}
    \left[\epsilon_{\alpha}  -\sigma g\mu B_{\alpha}/2 \right]
      d^{\dagger}_{\alpha\sigma} d_{\alpha\sigma}
        + \sum\limits_{\alpha} U_{\alpha}
           d^{\dagger}_{\alpha \uparrow}  d_{\alpha \uparrow}
              d^{\dagger}_{\alpha \downarrow}  d_{\alpha \downarrow}
              \nonumber\\
& & + \sum\limits_{k\sigma\alpha}
    \epsilon_{\alpha k}
        a^{\dagger}_{\alpha k\sigma} a_{\alpha k\sigma}
    +  \sum\limits_{k\sigma\alpha} \left[ t_{\alpha k}
        a^{\dagger}_{\alpha k\sigma} d_{\alpha\sigma}
          + H.c. \right]  \nonumber \\
& & +\sum\limits_{\sigma}
    \left[ t_C e^{i\int^t_0dt' \Delta cos\omega t'}
    d^{\dagger}_{L\sigma} d_{R\sigma} +H.c.
                \right],
\label{H2}
\end{eqnarray}
where $\Delta\equiv \Delta_L-\Delta_R$. In (\ref{H2}), we take the
last term which explicitly depends on time $t$ as the interacting
part $H_I$, and the remaining part as $H_o\equiv H-H_I$. The
Green's function of $H_o$, $g^r(\epsilon)$, can be easily obtained
with a decoupling approximation at the Hartree level\cite{foot3},
\begin{equation}
g^r_{\alpha\alpha\sigma}(\epsilon) =
 \frac{\epsilon^-_{\alpha\sigma}
         +U_{\alpha}n_{\alpha\bar{\sigma}} }
      {(\epsilon-\epsilon_{\alpha\sigma})
           \epsilon^-_{\alpha\sigma}
          +\frac{i}{2} \Gamma_{\alpha}
           (\epsilon^-_{\alpha\sigma}
            +U_{\alpha}n_{\alpha\bar{\sigma}} ) },
\label{gr1}
\end{equation}
where $\epsilon^-_{\alpha\sigma}\equiv
\epsilon-\epsilon_{\alpha\sigma}-U_{\alpha}$,
$\epsilon_{\alpha\sigma}\equiv\epsilon_{\alpha}-\sigma g\mu
B_{\alpha}/2$, and $n_{\alpha\bar{\sigma}}$ is the time-averaged
intradot electron occupation number at the state $\bar{\sigma}$ in
the $\alpha$-QD which we solve self-consistently. It is worth
mentioning that $g^r_{\alpha\alpha\sigma}(\epsilon)$ in
Eq.(\ref{gr1}) has two resonances: one is at
$\epsilon_{\alpha\sigma}$ while its associated state at
$\epsilon_{\alpha\bar{\sigma}}$ is empty; the other resonance is
at $\epsilon_{\alpha\sigma}+U_{\alpha}$ while its associated state
$\epsilon_{\alpha\bar{\sigma}}$ is occupied. Notice, in $H_o$ the
left part of the spin-cell ({\it i.e.} the left-lead and the
left-QD) is not coupled with the right part of the spin-cell,
therefore they are in equilibrium respectively. Hence the Keldysh
Green's function $g^<_{\alpha\alpha\sigma}(\epsilon)$ for $H_o$
can be solved from the fluctuation-dissipation theorem: $
g^<_{\alpha\alpha\sigma}(\epsilon)
 =-f_{\alpha} \left[ g^r_{\alpha\alpha\sigma}(\epsilon)
     -g^a_{\alpha\alpha\sigma}(\epsilon)\right].
$ With these preparations, the Green's function $G^r$ and $G^<$ of
the total Hamiltonian $H$ can be solved. In particular, we
calculate $ G^r_{\alpha\beta \sigma }(t,t') \equiv -i\theta(t-t')
<\{d_{\alpha\sigma}(t), d^{\dagger}_{\beta\sigma}(t')\}>$ by
iterating the Dyson equation. In Fourier space, the Dyson equation
can be reduced to\cite{ref6,add2}
$$
{\bf G}^r_{\sigma;mn}(\epsilon) =
 {\bf g}^r_{\sigma;mn}(\epsilon)  +
  \sum\limits_k {\bf G}^r_{\sigma;mk}(\epsilon)
{\bf \Sigma}^r_{\sigma;kn}(\epsilon)  {\bf
g}^r_{\sigma;nn}(\epsilon),
$$
where ${\bf G}^r_{\sigma,mn}(\epsilon)\equiv {\bf
G}^r_{\sigma,n-m} (\epsilon+m\omega)$, and the quantity
$G_n(\epsilon)$ is the Fourier expansion of $G(t,t')$.\cite{ref6}
The retarded self-energy ${\bf \Sigma}^r_{\sigma;kn}(\epsilon)$ is
the Fourier transform of ${\bf \Sigma}^r_{\sigma}(t_1,t_2)$ where
$\Sigma^r_{LR\sigma}(t_1,t_2)  = \Sigma^{r*}_{RL\sigma}(t_1,t_2) =
\delta(t_1-t_2) t_C exp[i\int^{t_1}_0 dt' \Delta cos \omega t']$,
and $\Sigma^r_{LL\sigma}=\Sigma^r_{RR\sigma}=0$. We obtain $
\Sigma^r_{LR\sigma,mn}(\epsilon) =
\Sigma^{r*}_{RL\sigma,nm}(\epsilon) = t_C
J_{n-m}(\frac{\Delta}{\omega})$. The Green's function ${\bf
g}^r_{\sigma;mn}(\epsilon)$ is:
$g^r_{\alpha\beta\sigma;mn}(\epsilon) =\delta_{\alpha\beta}
\delta_{mn} g^r_{\alpha\alpha\sigma}(\epsilon+m\omega)$. Then
${\bf G}^r_{\sigma;mn}(\epsilon)$ can be solved from the above
Dyson equation:\cite{add3}
$$
G^r_{\alpha\alpha\sigma,mn}(\epsilon) =
 \delta_{mn} \left/\left[
 { \left(g^r_{\alpha\alpha\sigma,mm}\right)^{-1} -A_{\bar{\alpha}\sigma,mm} }
 \right]\right. ,
$$
$$
 G^r_{\alpha\bar{\alpha}\sigma,mn}(\epsilon) =
   G^r_{\alpha\alpha\sigma,mm}(\epsilon)
   \Sigma^r_{\alpha\bar{\alpha}\sigma,mn}
       g^r_{\bar{\alpha}\bar{\alpha}\sigma,nn}(\epsilon),
$$
where $ A_{\alpha\sigma,mm}(\epsilon) \equiv \sum\limits_k |t_C|^2
   J^2_{k-m}\left(\frac{\Delta}{\omega}\right)
      g^r_{\alpha\alpha\sigma;kk}(\epsilon) .
$ Afterwards, the total Keldysh Green's function $
G^<_{\alpha\beta \sigma }(t,t') \equiv
   i< d^{\dagger}_{\beta\sigma}(t') d_{\alpha\sigma}(t)>
$ is easily obtained from the Keldysh equation. Finally,  we
obtain the time-averaged current in lead $\alpha$ from
\begin{eqnarray}
I_{\alpha\sigma}& \equiv & <I_{\alpha\sigma}(t)> =
   -Im \int (d\epsilon/2\pi) \Gamma_{\alpha}(\epsilon) \left[
    G^<_{\alpha\alpha\sigma,00}(\epsilon) \right. \nonumber \\
& + & \left. 2f_{\alpha}(\epsilon)
     G^r_{\alpha\alpha\sigma,00}(\epsilon) \right] ,
\label{I1}
\end{eqnarray}
and the self-consistent equation for the intradot occupation
number $n_{\alpha\sigma}$: $n_{\alpha\sigma} =-i\int
(d\epsilon/2\pi)
  G^<_{\alpha\alpha\sigma,00}(\epsilon) $.

Fig.(2) shows the calculated charge current $I_e$ (in units of
$e$) and the spin current $I_s$ (units of $\hbar/2$) versus the
gate voltage $V_{gR}$ at different microwave frequency $\omega$.
$I_e$ shows a positive peak due to the $A+$ process and a negative
peak by the $A-$ process (see Fig.(1c)), but $I_s$ has two
positive peaks. As we tune the gate voltage $V_{gR}$, the right QD
level is shifted so that when
$\hbar\omega=\epsilon_{L\downarrow}+U_L-\epsilon_{R\downarrow}$,
the $A-$ process occurs with high probability leading to a
positive peak in $I_s$ and a negative peak in $I_e$. On the other
hand we get positive peaks in both $I_e$ and $I_s$ when
$\hbar\omega=\epsilon_{R\uparrow}+U_R-\epsilon_{L\uparrow}$, for
the the $A+$ process.

The peak positions in $I_e,I_s$ due to the $A\pm$ processes shift
linearly with the microwave frequency $\omega$, as shown by the
dotted lines in Fig.(2). Eventually, at a special frequency
indicated by $A$, {\it i.e.} when
$\hbar\omega^*=\epsilon_{R\uparrow}+U_R-\epsilon_{L\uparrow} =
\epsilon_{L\downarrow} +U_L-\epsilon_{R\downarrow}$, the two peaks
overlap so that the net charge current $I_e$ cancels exactly due
to the cancellation of the $A \pm$ processes, at the same time the
spin current $I_s$ doubles its value. At this special frequency,
the full operation of the spin-cell occurs so that a spin current
is driven across the spin-cell, from left electrode to the right
electrode, without a charge current. If we connect the spin-cell
to complete an external circuit, a spin current will be driven and
continue to flow across the spin-cell into the circuit.\cite{addnote} 
On the other hand, if we let the two poles
of the spin-cell open,  although $I_s$ must be zero, a spin-motive
force in the two poles of the spin cell will still be induced so
that chemical potential
$\mu_{\alpha\uparrow}\not=\mu_{\alpha\downarrow}$. For example, in
the case of Fig.(1c), an open circuit will lead to
$\mu_{L\uparrow} <\mu_{L\downarrow}$ and $\mu_{R\uparrow}
>\mu_{R\downarrow}$.

In the following we focus on the spin-cell operation by fixing
gate voltage $V_{gR}=0.45$ which is its value at point A of
Fig.(2). We investigate $I_e,I_s$ as functions of the overall gate
potential $V_g$ (Fig.(3a)), magnetic field $g\mu B_L$ (Fig.(3b)),
and frequency $\omega$ (Fig.(3c)). Different curves in Fig.(3)
correspond to different microwave strength $\Delta\equiv
\Delta_L-\Delta_R$. In all situations $I_e\approx 0$ and we do not
discuss it anymore. Fig.(3c) shows that $I_s$ has several peaks
and dips when we vary $\omega$: the large peak indicated by $A$ is
the spin-cell operation discussed above, but peaks at $C$ and $D$
correspond to double- and triple-photon processes which connect
the $A\pm$ transitions of Fig.(1c). The dip at $B$ originates from
less probable transitions connecting levels indicated by the
dashed lines of Fig.(1c), while the dip at $E$ is its two-photon
process. Now, fixing $\omega$ at $\omega^*$, {\it i.e.} at the
spin-cell operation point A, the value of $I_s$ can be tuned by
the overall gate voltage $V_g$ as shown in Fig.(3a). However,
$I_s$ keeps large values for a wide range of $V_g$: this range is
in the Coulomb interaction scale $U/e$! This is important, because
in an experimental situation any background charge or
environmental effect near the spin-cell may alter the overall
potential, and Fig.(3a) shows that the spin-cell operation is not
critically altered by this effect. When $V_g$ becomes very large
so that $\epsilon_{L\downarrow}+U_L$ and
$\epsilon_{R\uparrow}+U_R$ is below the chemical potential $\mu$,
or $\epsilon_{L\uparrow}$ and $\epsilon_{R\downarrow}$ is above
$\mu$, $I_s$ diminishes because the $A\pm$ processes can no longer
occur (see Fig.(1c)). Finally, a very important result is shown in
Fig.(3b), where we fixed $g\mu B_R=-0.4$ while varying $g\mu B_L$
at the spin-cell operation point\cite{foot1} $A$. Fig.(3b) shows
clearly that $I_s$ increases with an increasing difference of
($B_L-B_R$): $I_s=0$ identically when $B_L=B_R$ if $U_L=U_R$, or
$I_s\approx 0$ if $U_L\not=U_R$. However, Fig.(3b) demonstrates
that we only need a slight difference in $B_L$ and $B_R$, at a
scale of the coupling constant $\Gamma_{\alpha}$, to generate a
substantial $I_s$. The most important fact is that $B_L$ and $B_R$
do not have to point to opposite directions which is
experimentally difficult to do.  In fact, if the two QD's are
fabricated with different materials so that the $g-$factors are
different, one can actually use an {\it uniform} magnetic field
throughout.

The proposed spin-cell for spin current should be experimentally
feasible using present technologies. First, the double-QD
structures can and have been fabricated by several laboratories.
Second, microwave assisted quantum transport measurements have
recently been reported\cite{ref7,ref8,ref9}. In particular, the
asymmetrical microwave radiation on double-QD device ({\bf i.e.}
$\Delta_L\not=\Delta_R$) has already been carried out
experimentally\cite{ref7}. Third, the asymmetric magnetic field
should be feasible as we have discussed above. If one takes
$f=\omega/2\pi=50$GHz, arranges the corresponding
$U$($\sim\hbar\omega$)$\approx 0.2 meV$, and fixes the temperature
scale $K_B T$ and coupling $\Gamma_{\alpha}$ to be twenty times
less than $U$ as in typical QD experiments, {\i.e.} $k_B T=100mK$
and $\Gamma=10\mu eV$, the corresponding magnetic field difference
is ($g\mu(B_L-B_R)\sim \Gamma$) $|B_L-B_R| \sim 0.16/g$ tesla.
These QD parameters have already been realized by present
technology. Finally, it is not difficult to show that by adjusting
the gate voltages one can easily calibrate the spin-cell operating
point\cite{foot2}.

{\bf Acknowledgments:} We gratefully acknowledge financial support
from NSERC of Canada, FCAR of Quebec (Q.S., H.G), and a RGC grant
from the SAR Government of Hong Kong under grant number HKU
7091/01P (J.W.). H.G. thanks Dr. Junren Shi for a discussion on
photon assisted tunneling. We gratefully acknowledge Dr. Baigeng
Wang for many discussions and inputs on the physics of spin
current.


\begin{center}
{\bf Appendix A}
\end{center}

In this appendix, we give a detailed derivation process of the
unitary transformation in which the Hamiltonian (1) is transformed to
Hamiltonian (2) in the manuscript. We take the unitary matrix $U(t)$ 
as:
$$
U(t) = exp\left\{ i\int^t_0 dt' \sum\limits_{\alpha}
\left[W_{\alpha}(t') \left( \sum\limits_{k\sigma}
a^{\dagger}_{\alpha k\sigma} a_{\alpha k \sigma}
 + \sum\limits_{\sigma } d^{\dagger}_{\alpha\sigma} d_{\alpha\sigma}
\right)\right]\right\} \eqno{(A1)}
$$
Under this unitary transformation, the operator $X_{\alpha}$
($X_{\alpha}$ represents $a_{\alpha k\sigma}$ and $d_{\alpha
\sigma}$ ) and $X^{\dagger}_{\alpha}$ transform into:
$$
U(t) X_{\alpha} U^{\dagger}(t) = X_{\alpha} exp\left[-i\int^t_0 dt'
W_{\alpha}(t')\right] \eqno{(A2)}
$$
$$
U(t) X^{\dagger}_{\alpha} U^{\dagger}(t) = X^{\dagger}_{\alpha}
exp\left[i\int^t_0 dt' W_{\alpha}(t')\right] \eqno{(A3)}
$$
Because unitary matrix $U(t)$ is dependent of time $t$, the Hamiltonian 
$H$ after the unitary transformation is:
$$
H_{new}(t) = U(t) H(t) U^{\dagger}(t) - U(t)
i\frac{\partial}{\partial t} U^{\dagger}(t) \eqno{(A4)}
$$
Using Eqs.(A2) and (A3), we easily obtain the new Hamiltonian (2) as 
given 
in the manuscript. We emphasize that this transform is done exactly and 
no approximation is involved. 

\begin{center}
{\bf Appendix B}
\end{center}

In this Appendix, we present a detailed derivation for the retarded 
Green's 
function $g^r_{\alpha\alpha \sigma}(\epsilon)$ (Eq.(3) in the 
manuscript) 
of the Hamiltonian $H_0$. Here, we solve 
$g^r_{\alpha\alpha\sigma}(\epsilon)$ 
by using the equation of motion method:
$$
\epsilon <<\hat{A} | \hat{B} >>^r = <\{ \hat{A}, \hat{B} \}> + <<
[\hat{A}, H_0] | \hat{B} >>^r \eqno{(B1)}
$$
where $\hat{A}$ and $\hat{B}$ represent arbitrary operators, and
$<<\hat{A}|\hat{B}>>^r$ is the Fourier expansion of the retarded
Green's function $-i \theta(t) <\{ \hat{A}(t), \hat{B}(0) \}>$.
Because $H_0$ is independent of time $t$, 
$<<\hat{A}(t) |\hat{B}(t') >>^r$ only depends on the time difference
$t-t'$. Appling equation of motion to the Green function
$g^r_{\alpha\alpha\sigma}=<<d_{\alpha\sigma}|d^{\dagger}_{\alpha\sigma}>>^r$,
we have:
$$
(\epsilon-\epsilon_{\alpha\sigma})
<<d_{\alpha\sigma}|d^{\dagger}_{\alpha\sigma}>>^r =1
 + \sum\limits_{k} t^*_{k\alpha} 
<< a_{\alpha k\sigma} |d^{\dagger}_{\alpha \sigma} >>^r
 + U_{\alpha} <<d_{\alpha\sigma} d^{\dagger}_{\alpha \bar{\sigma}}
  d_{\alpha\bar{\sigma}} |d^{\dagger}_{\alpha\sigma }>>^r
\eqno{(B2)}
$$
and
$$
<< a_{\alpha k\sigma} | d^{\dagger}_{\alpha\sigma} >>^r =
 \frac{t_{k\alpha}}{\epsilon-\epsilon_{k\sigma} +i0^+}
 << d_{\alpha \sigma }| d^{\dagger}_{\alpha\sigma} >>^r
\eqno{(B3)}
$$
Substitute Eq.(B3) into Eq.(B2), we obtain:
$$
(\epsilon-\epsilon_{\alpha\sigma} -\Sigma^r_{\alpha})
<<d_{\alpha\sigma}|d^{\dagger}_{\alpha\sigma}>>^r =1 +U_{\alpha}
<<d_{\alpha\sigma} d^{\dagger}_{\alpha \bar{\sigma}}
  d_{\alpha\bar{\sigma}} |d^{\dagger}_{\alpha\sigma} >>^r
\eqno{(B4)}
$$
where $\Sigma^r_{\alpha} \equiv \sum\limits_k \frac
{|t_{k\alpha}|^2} {\epsilon-\epsilon_{k\sigma}+i0^+} =\frac{-i}{2}
\Gamma_{\alpha} $ is the self-energy function. 

At this point, if one takes a decoupling approximation for the new 
Green's 
function $<<d_{\alpha\sigma} d^{\dagger}_{\alpha \bar{\sigma}}
d_{\alpha\bar{\sigma}} |d^{\dagger}_{\alpha\sigma }>>^r$ in
Eq.(B4), {\it i.e.},
$$
<<d_{\alpha\sigma} d^{\dagger}_{\alpha \bar{\sigma}}
  d_{\alpha\bar{\sigma}} |d^{\dagger}_{\alpha\sigma }>>^r
= n_{\alpha \bar{\sigma}} <<d_{\alpha\sigma}
|d^{\dagger}_{\alpha\sigma} >>^r \eqno{(B5)}
$$
This is the mean field approximation. 

Rather, we continue apply the equation of motion to
$<<d_{\alpha\sigma} d^{\dagger}_{\alpha \bar{\sigma}} 
d_{\alpha\bar{\sigma}} |d^{\dagger}_{\alpha\sigma } >>^r$, and we have:
$$
(\epsilon-\epsilon_{\alpha\sigma}-U) <<d_{\alpha\sigma}
d^{\dagger}_{\alpha \bar{\sigma}} d_{\alpha\bar{\sigma}}
|d^{\dagger}_{\alpha\sigma }>>^r  = n_{\alpha\bar{\sigma}}
\hspace{60mm}
$$
$$
+ \sum\limits_k t^*_{\alpha k} <<a_{\alpha k\sigma}
d^{\dagger}_{\alpha \bar{\sigma}} d_{\alpha\bar{\sigma}}
|d^{\dagger}_{\alpha\sigma }>>^r
$$
$$
-\sum\limits_k t_{\alpha k} <<d_{\alpha\sigma} a^{\dagger}_{\alpha
k\bar{\sigma}} d_{\alpha\bar{\sigma}} |d^{\dagger}_{\alpha\sigma}
>>^r
$$
$$
+\sum\limits_k t^*_{\alpha k} <<d_{\alpha\sigma}
d^{\dagger}_{\alpha \bar{\sigma}} a_{\alpha k\bar{\sigma}}
|d^{\dagger}_{\alpha\sigma }>>^r \eqno{(B6)}
$$
In this equation, three new Green's functions $<<a_{\alpha k\sigma}
d^{\dagger}_{\alpha \bar{\sigma}} d_{\alpha\bar{\sigma}}
|d^{\dagger}_{\alpha\sigma }>>^r$, $<<d_{\alpha\sigma}
a^{\dagger}_{\alpha k\bar{\sigma}} d_{\alpha\bar{\sigma}}
|d^{\dagger}_{\alpha\sigma }>>^r$, and $<<d_{\alpha\sigma}
d^{\dagger}_{\alpha \bar{\sigma}} a_{\alpha k\bar{\sigma}}
|d^{\dagger}_{\alpha\sigma} >>^r$, emerge. We now take
the decoupling approximation:
$$
<<a_{\alpha k\sigma} d^{\dagger}_{\alpha \bar{\sigma}}
d_{\alpha\bar{\sigma}} |d^{\dagger}_{\alpha\sigma} >>^r =n_{\alpha
\bar{\sigma}} <<a_{\alpha k\sigma }| d^{\dagger}_{\alpha\sigma}
>>^r
$$
$$
<<d_{\alpha\sigma} a^{\dagger}_{\alpha k\bar{\sigma}}
d_{\alpha\bar{\sigma}} |d^{\dagger}_{\alpha\sigma} >>^r=
<<d_{\alpha\sigma} d^{\dagger}_{\alpha \bar{\sigma}} a_{\alpha
k\bar{\sigma}} |d^{\dagger}_{\alpha\sigma }>>^r=0 \eqno{(B7)}
$$
Eq.(B6) reduces into:
$$
(\epsilon-\epsilon_{\alpha\sigma}-U) <<d_{\alpha\sigma}
d^{\dagger}_{\alpha \bar{\sigma}} d_{\alpha\bar{\sigma}}
|d^{\dagger}_{\alpha\sigma }>>^r  = n_{\alpha\bar{\sigma}}
+\Sigma_{\alpha} n_{\alpha \bar{\sigma}} << d_{\alpha \sigma} |
d^{\dagger}_{\alpha\sigma} >>^r \eqno{(B8)}
$$
Then from Eqs.(B4) and (B8), we can solve the Green's function
$g^r_{\alpha\alpha\sigma} (\epsilon) = <<d_{\alpha\sigma}
|d^{\dagger}_{\alpha\sigma}>>^r$ as:
$$
g^r_{\alpha\alpha\sigma}(\epsilon) =
\frac{\epsilon^-_{\alpha\sigma} + U_{\alpha}
n_{\alpha\bar{\sigma}} } {(\epsilon-\epsilon_{\alpha\sigma})
\epsilon^-_{\alpha\sigma} -\Sigma^r_{\alpha}
(\epsilon^-_{\alpha\sigma}
 + U_{\alpha} n_{\alpha\bar{\sigma}} ) }
\eqno{(B9)}
$$
This is the Eq.(3) in the manuscript.

If one applies the equation of motion one more time, on the three new
Green's functions of Eq.(B6), then takes a decoupling approximation, 
the Green function is obtained to be:
$$
g^r_{\alpha\alpha\sigma}(\epsilon) = \frac{1+U_{\alpha}
A_{\alpha\sigma} n_{\alpha\bar{\sigma}} } {
\epsilon-\epsilon_{\alpha\sigma}+\frac{i}{2} \Gamma_{\alpha}
 +U_{\alpha} A_{\alpha\sigma} (\Sigma^{(a)}_{\alpha\bar{\sigma}}
+\Sigma^{(b)}_{\alpha\bar{\sigma}} ) } \eqno{(B10)}
$$
where the definition of $A_{\alpha\sigma}$,
$\Sigma^{(a)}_{\alpha\bar{\sigma}}$ and
$\Sigma^{(b)}_{\alpha\bar{\sigma}}$, are given in Phys. Rev. B
64,153306 (2001). This solution has a Kondo resonance and it has
been applied in some previous work for investigating the Kondo
effect.

\begin{center}
{\bf Appendix C}
\end{center}

In this Appendix, we give a detailed derivation for
the Green's function $G^r_{\alpha\beta\sigma,mn}(\epsilon)$ of the
Hamiltonian $H$. First, from the Dyson equation, we have:
$$
G^r_{LL\sigma,mn} =g^r_{LL\sigma,mn} + \sum\limits_k
G^r_{LR\sigma,mk} \Sigma^r_{RL\sigma,kn} g^r_{LL\sigma,nn}
\eqno{(C1)}
$$
$$
G^r_{LR\sigma,mn} =\sum\limits_k G^r_{LL\sigma,mk}
\Sigma^r_{LR\sigma,kn} g^r_{RR\sigma,nn} \eqno{(C2)}
$$
where we have suppressed the argument $\epsilon$. Substitute
Eq.(C2) into Eq.(C1), we obtain:
$$
G^r_{LL\sigma,mn} = g^r_{LL\sigma,mn} +
\sum\limits_{k_1,k_2} G^r_{LL\sigma,mk_1} \Sigma^r_{LR\sigma,k_1k_2}
g^r_{RR\sigma,k_2 k_2} \Sigma^r_{RL\sigma,k_2 n }
g^r_{LL\sigma,n n} \eqno{(C3)}
$$
Introducing notation $A_{R\sigma,mn}(\epsilon)$,
$$
A_{R\sigma,mn}(\epsilon) \equiv \sum\limits_k \Sigma^r_{LR\sigma,mk}
g^r_{RR\sigma,kk} \Sigma^r_{RL\sigma,kn} \eqno{(C4)}
$$
then the Eq.(C3) reduces to:
$$
G^r_{LL\sigma,mn} = g^r_{LL\sigma,mn} +
\sum\limits_{k_1} G^r_{LL\sigma,mk_1} A_{R\sigma,k_1 n}
g^r_{LL\sigma,n n}
$$
$$
=g^r_{LL\sigma,mn} + \sum\limits_{k_1}
g^r_{LL\sigma,mk_1} A_{R\sigma,k_1 n} g^r_{LL\sigma,n n}
$$
$$
 + \sum\limits_{k_1 k_2} g^r_{LL\sigma,mk_1} A_{R\sigma,k_1 k_2}
g^r_{LL\sigma,k_2 k_2} A_{R\sigma, k_2 n} g^r_{LL\sigma,n n} +
... \eqno{(C5)}
$$

In the following we take an approximation as our previous work [Eq.(29)
in Phys. Rev B 59,13126 (1999)]:
$$
A_{R\sigma,k_1 n}  g^r_{LL\sigma,k_1 k_1}  g^r_{LL\sigma,n n}
=A_{R\sigma, k_1 n} \delta_{k_1 n} \left[g^r_{LL\sigma,n n}\right]^2
\eqno{(C6)}
$$
This approximation is reasonable when $\hbar\omega >> max
(\Gamma_{\alpha} ,t_C)$.
In the following we discuss the physical picture of this
approximation. The Green's function $G^r_{LL\sigma, mn}$ is a
progapator (of the Hamiltonian $H$) for an electron from the left
dot to the left dot. The first term $g^r_{LL\sigma, mn}$ 
describes that the electron always stays in the left dot. The second 
term $\sum\limits_{k_1} g^r_{LL\sigma,mk_1} A_{R\sigma,k_1 n} 
g^r_{LL\sigma,n n}$ describes the progapation process in which the 
electron from left dot transits to the right dot while absorbing $k_1-k$ 
photons, it then stays in the right dot for some time, then goes back 
to
the left dot emitting $n-k$ photons. The higher order term describes 
multi-time transit processes between the left dot and the right dot. In 
our
approximation ({\it e.g.} for the 2nd term), we take $k_1=n$ and 
neglect
others. We emphasize that this term is indeed the most important.
For example, if $k_1 \not= n$, we consider an initial electron in the 
left dot 
sitting at the resonance state ({\it e.g.} at $\epsilon_{L\uparrow}$).
This electron tunnels to the right dot absorbing $k_1-k$ photons and goes 
back to the left dot emitting $n -k$ photons. Then this electron has 
energy $\epsilon_{L\uparrow} +(k_1 -n)\hbar\omega$, which has a 
difference 
$(k_1 -n)\hbar\omega$ from the resonance state $\epsilon_{L\uparrow}$, 
and 
there is no state at energy $\epsilon_{L\uparrow} +(k_1 -n)\hbar\omega$.
Therefore this processes ($k_1\not=n$) has very small probability to 
occur. 
In fact, when $\epsilon$ is near a resonant state (e.g. $\epsilon_{L\uparrow}$),
$g^r_{\alpha\alpha\sigma}(\epsilon)$ [Eq.(3) in the manuscript] can be rewritten as
$1/(\epsilon-\epsilon_{L\uparrow} +i\delta)$. Then 
the term in the left side of Eq.(C6) can be rewritten as:
$$
\frac{A_{R\sigma,k_1 n}} {(\epsilon -\epsilon_{L\uparrow}
+k_1\hbar\omega +i\delta) (\epsilon -\epsilon_{L\uparrow}
+n\hbar\omega +i\delta) }\eqno{(C7)}
$$
where $\delta $ is a small positive real number ($\sim \Gamma$).
Now it clearly shows that when $k_1 \not= n$, this term is always very 
small for any electron energy $\epsilon$ [note that
$A_{R\sigma,k_1 n}$ has a small factor $t^2_C$]. When $k_1 =n$,
this term is large at $\epsilon =\epsilon_{L\uparrow} -
n\hbar\omega $. Therefore our approximation is reasonable
if $\hbar\omega >> max (\Gamma_{\alpha}, t_C)$.

Under this approximation, we can easily solve $G^r_{LL\sigma,mn}$ from
Eq.(C5)
$$
G^r_{LL \sigma,mn}(\epsilon) = \frac{\delta_{mn}}
{\left(g^r_{LL\sigma,mm}\right)^{-1} -A_{R\sigma,mm} } \eqno{(C8)}
$$
Then,  $G^r_{LR\sigma,mn}$ can be obtained immediately:
$$
G^r_{LR\sigma,mn}(\epsilon) = G^r_{LL\sigma,mm}(\epsilon)
\Sigma^r_{LR\sigma,mn} g^r_{RR\sigma,nn}(\epsilon) \eqno{(C9)}
$$
$G^r_{RR\sigma,mn}(\epsilon)$ and
$G^r_{RL\sigma,mn}(\epsilon)$ can also be solved in the same manner.

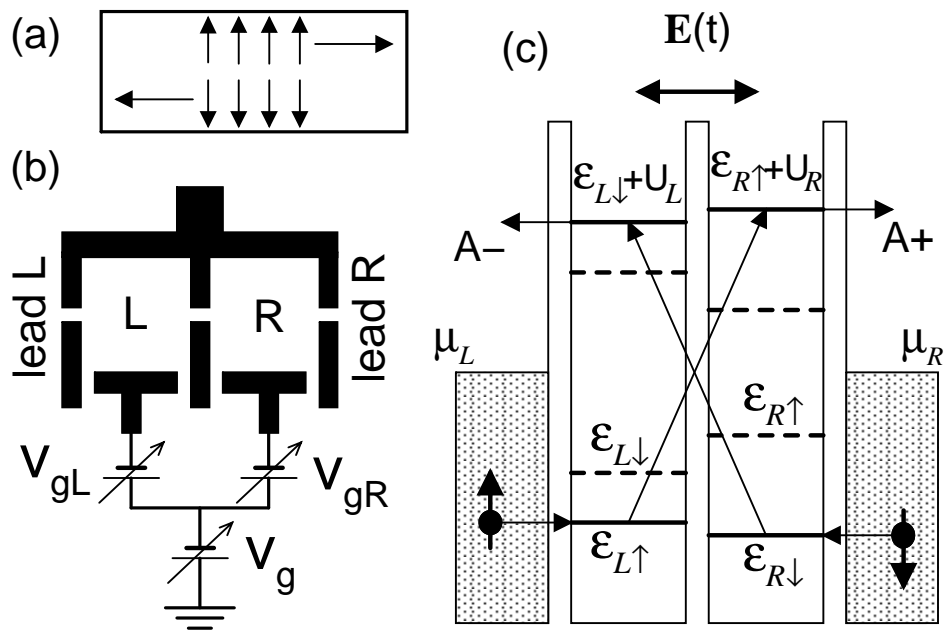
\begin{figure}
\caption{ (a). Schematic diagram for a conductor which has a spin
current with zero charge current; (b). Schematic diagram for the
double quantum dot spin-cell; (c). Schematic plot for the
spin-cell operation via photon-assisted tunneling processes
indicated by $A\pm$. } \label{fig1}
\end{figure}

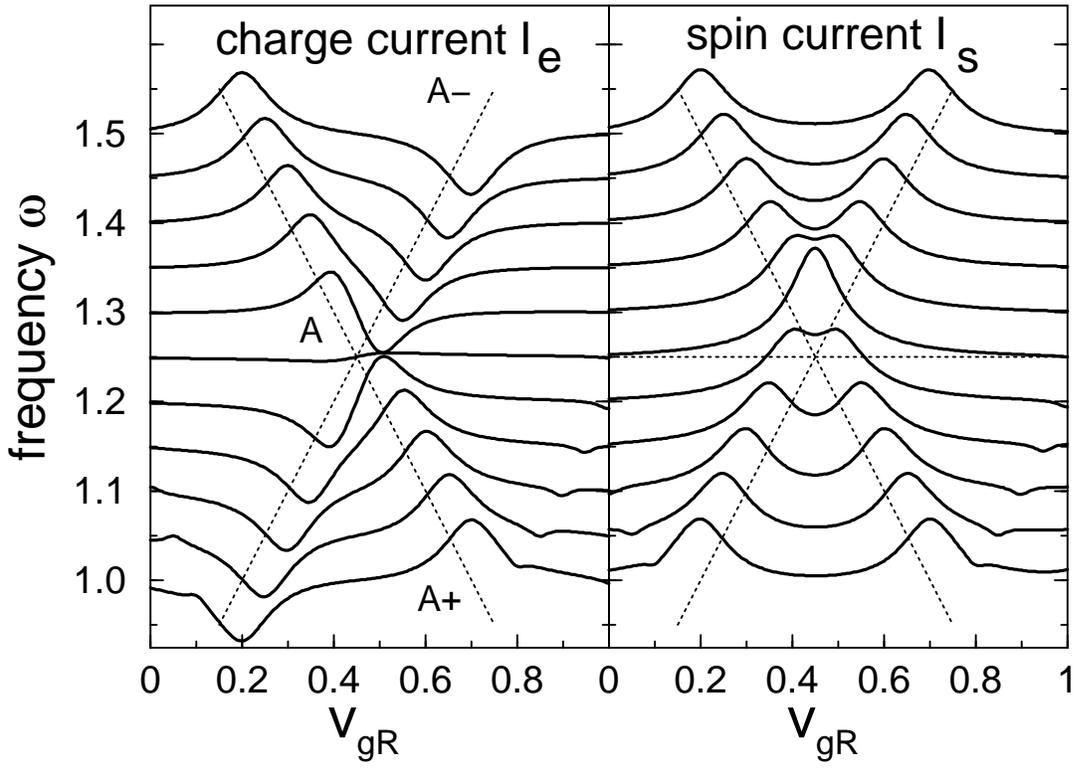
\begin{figure}
\caption{ The charge current $I_e$ and spin current $I_s$ versus
gate voltage $V_{gR}$ for different frequencies $\omega$.
Different curves have been offset such that the vertical axis
gives the frequency. Two dotted oblique lines $A\pm$ indicate the
position of the peaks. The parameters are: $\mu_L=\mu_R=0$,
$\Gamma_L=\Gamma_R=k_BT=0.1$, $t_C=0.02$, $U_L=1$, $U_R=0.9$,
$g\mu B_L=0.2$, $g\mu B_R=-0.4$, $V_{gL}=0.5$, $V_g=0$, and
$\Delta/\omega=1.0$. } \label{fig2}
\end{figure}

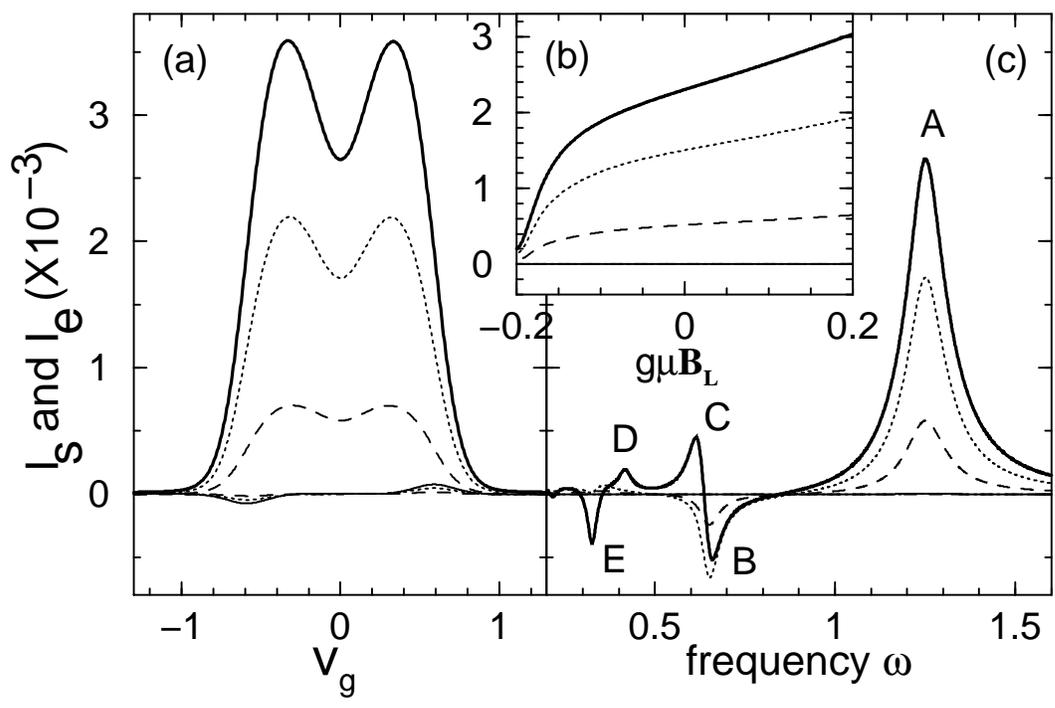
\begin{figure}
\caption{ (a), (b), and (c) are $I_e$ and $I_s$ versus gate
voltage $V_{g}$, the magnetic field $g\mu B_L$, and frequency
$\omega$, respectively. $\omega=1.25$ in (a); $2\omega=U_L+U_R
+g\mu (B_L-B_R)$ in (b). $V_{gR}=0.45$, and other parameters are
the same as Fig.2. The solid, dotted, and dashed lines correspond
to $\Delta/\omega =2.0$, $1.0$, and $0.5$, respectively. Notice
that the three curves of charge current overlap and they are all
essentially zero. } \label{fig3}
\end{figure}


\begin{references}

\bibitem{ref1}
S.A. Wolf, {\sl et al}, Science {\bf 294}, 1488 (2001).

\bibitem{ref2}
G.A. Prinz, Science {\bf 282}, 1660 (1998).

\bibitem{ref3}
There have been some work on spin current generation by a rotating
magnetic field in a {\it uni-pole} device. A uni-pole system,
however, cannot function as a spin-cell because it cannot complete
a spin-circuit. See, A. Brataas, Y. Tserkovnyak, G.E.W. Bauer, and
B. Halperin, Phys. Rev. B {\bf 66}, 060404 (2002); B. Wang, J.
Wang, and H. Guo, cond-mat/0208475 (2002).

\bibitem{ref4}
A.-P. Jauho, N.S. Wingreen, and Y. Meir, Phys. Rev. B {\bf 50},
5528 (1994).

\bibitem{ref7}
T.H. Oosterkamp, {\sl et al}, Nature {\bf 395}, 873 (1998).

\bibitem{ref5}
N.S. Wingreen, A.-P. Jauho, and Y. Meir, Phys. Rev. B, {\bf
48},8487 (1993).

\bibitem{foot3}
In deriving the (nonperturbed) retarded Green's function of $H_o$,
we have taken a decoupling approximation as:
\begin{eqnarray}
\ll a_{\alpha k\sigma} d^{\dagger}_{\alpha\bar{\sigma}}
    d_{\alpha\bar{\sigma}} | d^{\dagger}_{\alpha\sigma} \gg^r
    & =& n_{\alpha\bar{\sigma}} \ll a_{\alpha k\sigma}
                | d^{\dagger}_{\alpha\sigma} \gg^r ,
                        \nonumber \\
\ll a^{\dagger}_{\alpha k\bar{\sigma}} d_{\alpha\sigma}
    d_{\alpha\bar{\sigma}} | d^{\dagger}_{\alpha\sigma} \gg^r
            & = &
\ll a_{\alpha k\bar{\sigma}} d^{\dagger}_{\alpha\bar{\sigma}}
                     d_{\alpha\sigma} | d^{\dagger}_{\alpha\sigma} \gg^r
                              = 0 .
             \nonumber
\end{eqnarray}
In this approximation the level renormalization has been neglected. Because 
our system is in the Coulomb blockade regime, the level renormalization 
is very small and this approximation is reasonable. Even if the 
level renormalization is included, it does not affect the working principle
of the spin-cell.

\bibitem{ref6}
Q.-f. Sun, J. Wang, and T.-h. Lin, Phys. Rev. B {\bf 59}, 13126
(1999).

\bibitem{add2}
Because $H_0$ has interaction $U_{\alpha}$, this Dyson equation is not 
exact, but is a good approximation.

\bibitem{add3}
Here we took the same approximation as that of Ref. \onlinecite{ref6} which
is reasonable when $\hbar\omega >> max(\Gamma_{\alpha}, t_C)$.

\bibitem{addnote}
If resistences of the external circuit for spin-up and spin-down channels 
are slightly different, the spin-cell will drive a large spin current 
perhaps with a small charge current. However, by regulating the gate
voltage $V_g$ which makes the spin-motive force also slightly different
for spin-up and spin-down electrons, we can still obtain a spin-current with 
zero charge-current.

\bibitem{foot1}
The frequency $\omega^*$ of the spin-cell operation point $A$ (see
Fig.(1c)) actually depends on the value of $B_L$: $2\hbar\omega^*=
\epsilon_{L\downarrow}+U_L+\epsilon_{R\uparrow}+U_R
-\epsilon_{L\uparrow}-\epsilon_{R\downarrow} = U_L+U_R+g\mu
(B_L-B_R)$. To plot Fig.(3b) we varied $\omega^*$ for each value
of $B_L$ accordingly.

\bibitem{ref8}
T.H. Oosterkamp, {\sl et al}, Phys. Rev. Lett. {\bf 78}, 1536
(1997).

\bibitem{ref9}
L.P. Kouwenhoven, {\sl et al}, Phys. Rev. Lett. {\bf 73}, 3443
(1994).

\bibitem{foot2}
In order to calibrate experimental conditions at the spin-cell
operating point, one needs a method to detect spin current outside
the spin-cell. Recently, Hirsch has advanced a theoretical idea
for this purpose which works even in the absence of a charge
current: J.E. Hirsch, Phys. Rev. Lett. {\bf 83}, 1834 (1999).
Moreover, the detection can be made easier if we allow and then
detect a small charge current that flows through the spin-cell,
using the two panels of Fig.(2) as a ``map'' between the charge
and spin currents.

\end{references}
\end{document}